\documentclass[11pt]{emulateapj}
\usepackage{natbib}
\begin{document}
\newcommand{\hethree}{\ensuremath{^{3}\textrm{He}}}

\title{A \hethree{} driven instability near the fully convective boundary}
\shorttitle{}
\author{Jennifer L. van Saders and Marc H. Pinsonneault}
\affil{Department of Astronomy, The Ohio State University, 140 West 18th Avenue, Columbus, OH, 43210}
\email{vansaders@astronomy.ohio-state.edu,pinsono@astronomy.ohio-state.edu}
\slugcomment{submitted to ApJ}
\shortauthors{van Saders \& Pinsonneault}
\shorttitle{}
\begin{abstract}
We report on the discovery of an instability in low mass stars just above the threshold ($\sim 0.35 \textrm{ M}_{\odot}$) where they are expected to be fully convective on the main sequence. Non-equilibrium \hethree{} burning creates a convective core, which is separated from a deep convective envelope by a small radiative zone. The steady increase in central \hethree{} causes the core to grow until it touches the surface convection zone, which triggers fully convective episodes in what we call the ``convective kissing instability''. These episodes lower the central abundance and cause the star to return to a state in which is has a separate convective core and envelope.  These periodic events eventually cease when the \hethree{} abundance throughout the star is sufficiently high that the star is fully convective, and remains so for the rest of its main sequence lifetime. The episodes correspond to few percent changes in radius and luminosity, over Myr to Gyr timescales. We discuss the physics of the instability, as well as prospects for detecting its signatures in open clusters and wide binaries. Secondary stars in cataclysmic variables (CVs) will pass through this mass range, and this instability could be related to the observed paucity of such systems for periods between two and three hours. We demonstrate that the instability can be generated for CV secondaries with mass-loss rates of interest for such systems, and discuss potential implications.

\end{abstract}

\section{Introduction}

Stars are surprisingly stable for most of their evolution.  The exceptions provide windows into important transitions in stellar properties.  The helium flash, for example, marks the temporal transition from shell hydrogen burning to core helium burning.  Interiors theory also predicts changes in structure as a function of mass.  A classic example is the prediction that progressively lower mass models should have progressively deeper surface convection zones, culminating in a transition to a fully convective state at around a third of a solar mass.  The global behavior of models near this transition is traditionally thought to be smooth.

We have uncovered a novel instability for models just above the fully convective boundary that has, to the best of our knowledge, never before been noted for these objects.  Non-equilibrium \hethree{} burning in a marginally stable and small radiative core leads to a steady increase in \hethree{} and energy generation, and the development of a small convective core separated from the convective envelope by a radiative buffer zone.  Above a critical threshold a transition to a fully convective state occurs.  Convective mixing lowers the core \hethree{} abundance, quenching the nuclear reactions and causing the core to become radiative again.  The sudden drop in energy generation leads to a sudden drop in radius, followed by a gradual recovery as \hethree{} production in the core resumes.  This phenomenon repeats in a series of pulses, with an amplitude which is steadily damped as the envelope \hethree{} is enriched and the drop in core \hethree{} becomes smaller.  For non-interacting stars this will apply to a narrow range of masses just above the fully convective boundary.  The secondary stars of cataclysmic variables, however, will all pass through this mass range, and the onset of the instability occurs at the upper end of the CV period gap.  The astrophysical consequences may therefore be surprisingly broad, and the instability can be triggered for mass-loss rates of interest in the CV context.  In this paper we describe the physics of the instability and present some preliminary assessment of its effects.  We describe our stellar models in Section \ref{sec:models}, discuss the details of the physical processes responsible for the instability in Section \ref{sec:physics}, observational implications of the instability in Section \ref{sec:observations}, and close with our conclusions in Section \ref{sec:conclusion}.

\section{Stellar Models} \label{sec:models}
We utilize the Yale Rotating Stellar Evolution Code (YREC) \citep{pinsonneault1989, bahcall1992, bahcall1995, bahcall2001} to produce a grid of stellar models with a standard set of input physics. We create a dense grid near the fully convective boundary to investigate the mass range over which this instability operates. Models range in mass from $0.2-0.5 \textrm{ M}_{\odot}$ and are spaced every $0.001 \textrm{ M}_{\odot}$. Models generally contain $\sim 500$ shells and take $\sim 5000$ timesteps reach to 14 Gyr, at which point the evolution is truncated. All models include helium and heavy element diffusion following the procedure of \citet{thoul1994}, atmosphere and boundary conditions of \citet{allard2000}, nuclear reaction rates of \citet{adelberger2011} with weak screening \citep{salpeter1954}, and employ a mixing length theory of convection \citep{cox1968, vitense1953}. Opacities are from the Opacity Project (OP) \citep{mendoza2007} for a \citet{grevesse1998} solar mixture, supplemented with the low temperature opacities of \citet{ferguson2005}. We utilize the 2006 OPAL equation of state \citep{rogers1996,rogers2002} and the \citet{saumon1995} EOS for temperature and density combinations outside of the OPAL tables. Convective instability is determined using the Schwarzschild criterion. We impose an initial \hethree{} mass fraction of $2.95\times 10^{-5}$, for a \citet{grevesse1998} mixture with the $\textrm{He}^3 / \textrm{He}^4$ isotopic ratio from \citet{anders1989} on a series of pre-main sequence starting models. Because $pp$ chain burning is by far the most significant source of \hethree{}, the exact value of this initial abundance is of relatively little importance. 

\section{\hethree{} instability} \label{sec:physics}

We find that low mass objects at the fully convective boundary undergo non-equilibrium \hethree{} burning during the early stages of their main-sequence lifetimes which gives rise to periodic fully convective episodes. We discuss the nature of the instability, the conditions under which it operates, and compare our results obtained with YREC to those obtained with the MESA stellar evolution code. 

\begin{figure}
	\centerline{\includegraphics[scale = 1.0]{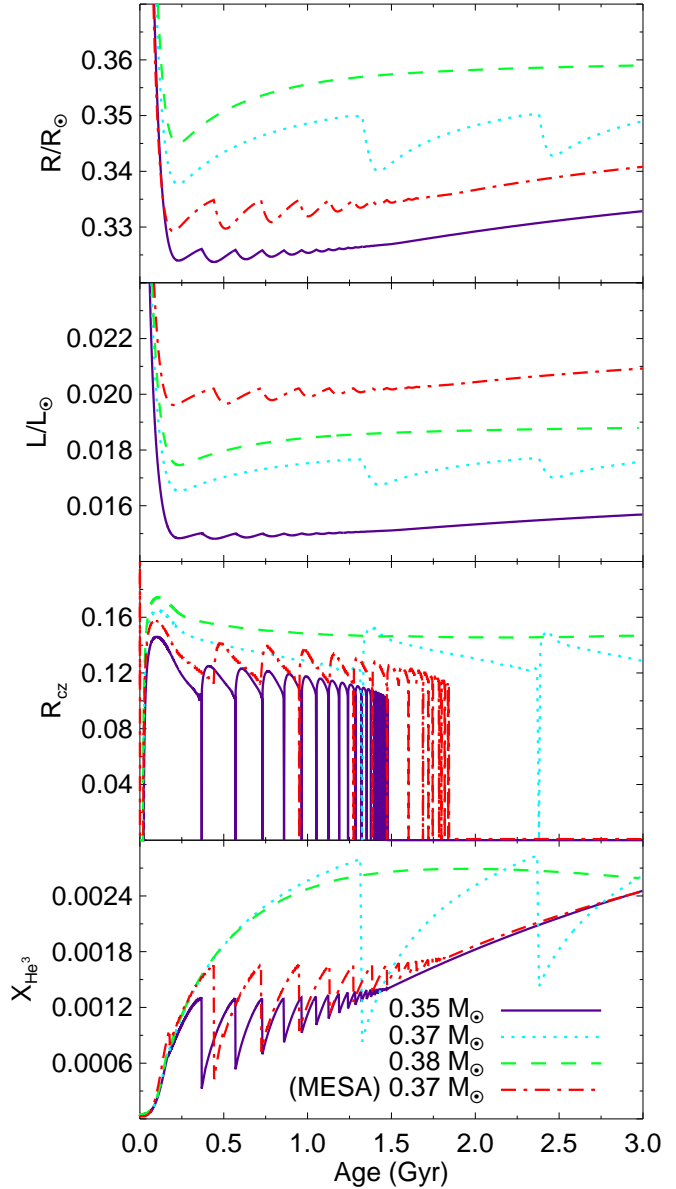}}
       \caption{Behavior of models of several characteristic masses through convective episodes. The panels show radius, luminosity, location of the base of the convective envelope, and core $\textrm{He}^3$ fraction from top panel to bottom panel in order, each as a function of time. Curves for YREC models are as follows: solid purple: $0.35 \textrm{ M}_{\odot}$ model, blue dotted: $0.37 \textrm{ M}_{\odot}$, green dashed: $0.38 \textrm{ M}_{\odot}$ (which is too massive to show convective episodes). The red dot-dashed curve shows a MESA $0.37 \textrm{ M}_{\odot}$ model. }
	\label{fig1}
\end{figure} 

\subsection{Nature of the Instability}

\hethree{} is produced in low mass stars via proton capture on deuterium; the former can be either primordial or produced from the $pp$ chain.  The $pp$ reaction itself has a low cross section but a weak temperature dependence, so \hethree{} can be generated even in relatively cool environments given sufficient time.  \hethree{} is destroyed by both \hethree{} + \hethree{} ($pp$ I) and \hethree{} + $^{4}\textrm{He}$ reactions ($pp$ II and $pp$ III); the former is far more important at low temperatures.   At high temperatures equilibrium is achieved with a \hethree{} abundance that decreases as the temperature increases, and net destruction of \hethree{} can occur.  However, at low temperatures the rate of production exceeds the rate of destruction and \hethree{} is produced (\citealt{iben1967};  \citealt{rood1976}; see also \citealt{boesgaard1985} for a discussion). One therefore expects a \hethree{} abundance peak in the outer layers of higher mass stars \citep[see Figure 1 in][for an example]{pinsonneault1989}. For lower mass stars the peak shifts towards the center of the star and the development of a deep convective envelope homogenizes the outer layers.  For sufficiently cool central temperatures \hethree{} is never in equilibrium.  Therefore the nuclear energy generation from \hethree{} + \hethree{} burning will depend sensitively on the integrated \hethree{} production over the lifetime of the star. 

 If there is a radiative core it is possible to have a local enhancement in the \hethree{} abundance which is much larger than the average for a fully convective star.  This special circumstance can therefore create a small convective core, even in stars far too cool to support the CNO burning responsible for convective cores on the upper main sequence.  The growth in central \hethree{} will cause this core to grow in mass.  For a sufficiently small radiative core it is therefore possible to induce a ``convective kissing'' instability where the central and surface convective regions merge.

\begin{figure}
	\centerline{\includegraphics[scale = 1.0]{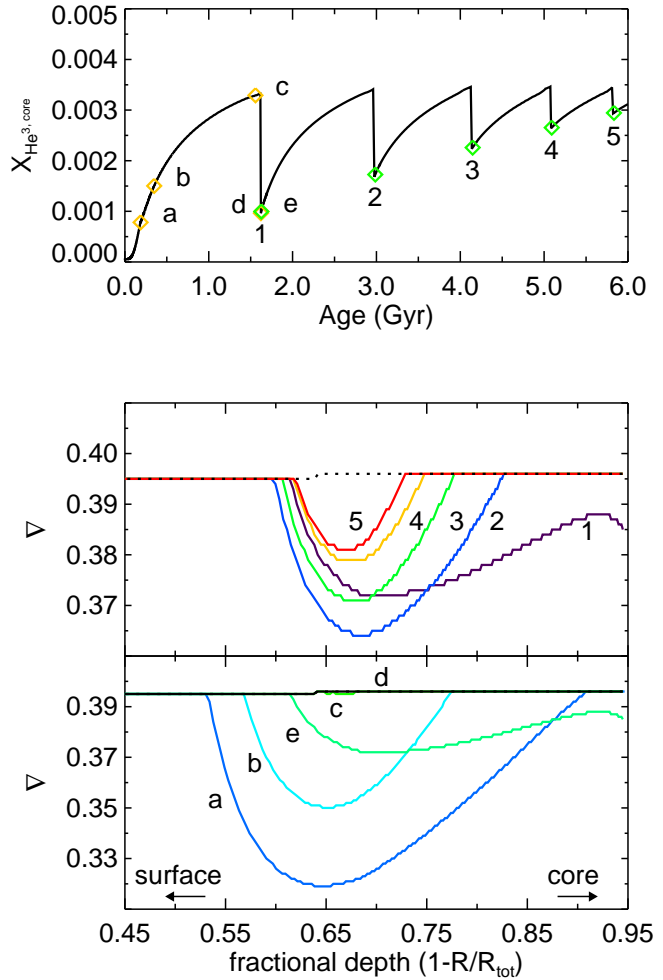}}
       \caption{Temperature gradients and core \hethree{} abundance in a $M = 0.360 \textrm{ M}_{\odot}$ model. The bottom two panels show the temperature gradients at different times during the evolution of the model, which are labeled on the top panel on a plot of the core \hethree{} abundance. In the bottom two panels, the adiabatic temperature gradient is plotted as a thick dashed black line. Any curves that fall below this line represent radiative regions, and any curves that follow it convection zones. }
	\label{fig2}
\end{figure}

To understand the nature of the this instability, we introduce Figure \ref{fig1}, which contains plots of the location of the base of the envelope convection zone, core \hethree{} fraction ($X_{\hethree{}}$), luminosity, and radius of several stellar models of different masses. Fully convective episodes (when $R_{cz} = 0$) are in phase with sudden drops in $X_{\hethree{}}$ in the core. After an episode the fraction of \hethree{} in the core slowly recovers, along with the luminosity from \hethree{} burning reactions, whose rates are inflated by the presence of the elevated \hethree{} abundances. The total luminosity and radius of the star both increase on nuclear timescales in response to the additional burning. The response of the core and envelope convection zones is shown in Figure \ref{fig2}, which plots the temperature gradients of a single $0.360 \textrm{ M}_{\odot}$ model at different stages in the convective cycles. At times early in the cycle ((a), bottom panel of Figure \ref{fig2}), the core and envelope convection zones are separated significantly in radius. As time passes and the mean \hethree{} of the entire star increases with each mixing episode, the increased \hethree{} burning inflates the radiative temperature gradient, resulting in a decrease in the separation between the envelope and core convection zones, as seen at times (b) and (c). By time (d) the additional luminosity is sufficient to make the convection zones meet, at which point the star is fully convective, and mixes the \hethree{} in the core throughout the entire model on a convective overturn timescale. In the case of a $0.36 \textrm{M}_{\odot}$ model, the luminosity from the reactions $^{3}\textrm{He} + ^{3}\textrm{He}$ and $^{3}\textrm{He} + ^{4}\textrm{He}$ in the convective core represents between $10$ and $35\%$ of the total model luminosity, depending on the location within the cycle. The dilution of the \hethree{} in the core results in decreased luminosity and thermal support, and the model radius and luminosity decrease on a thermal timescale. The sudden loss of substantial \hethree{} burning is somewhat counter-balanced by energy release due to gravitational collapse, which results in modest, few percent changes in the \emph{total} luminosity over the course of these episodes. The structure returns to a convective core, radiative buffer zone, and convective envelope, while the \hethree{} again begins to accumulate in the core.

 The middle panel of Figure \ref{fig2} shows the temperature gradient at the timestep immediately following the fully convective episodes for several cycles. Each successive cycle begins with higher \hethree{} abundances throughout the model, since the products of previous cycles are mixed throughout the star. Because the total \hethree{} abundance is higher at the onset than in previous cycles, the boundaries of the convective core and envelope begin the cycle more closely spaced in radius, and the duration of the cycles decreases. Cycles continue until the \hethree{} abundance throughout the star is high enough that the model is fully convective for the rest of its MS lifetime. 

Figure \ref{fig1} demonstrates the response of the convection zone, central \hethree{} abundance, radius and luminosity for models of different masses over the course of many convective cycles. More massive models tend to have cycles of longer duration and greater amplitude, because they are further from the fully convective boundary, and thus require more \hethree{} burning to cause them to be convectively unstable. Likewise, lower mass models have shorter cycles with smaller amplitude variations in core \hethree{} fraction, luminosity, and radius. In our models (at solar composition $X_i = 0.710342, Z_{i} = 0.018338$, with a solar calibrated mixing length parameter $\alpha = 1.924897 $), the instability occurs over the mass range $0.322 \textrm{ M}_{\odot}< M < 0.365 \textrm{ M}_{\odot}$. For masses $0.266 \textrm{ M}_{\odot}< M < 0.322 \textrm{ M}_{\odot}$ the star undergoes a brief period in which it has a radiative zone between a convective core and convective envelope without any episodic mixing, but is then fully convective on the remainder of the MS. Models with masses $M \le 0.266 \textrm{ M}_{\odot}$ are fully convective (in that they \emph{never} have a radiative core) throughout their MS lifetimes. This fully convective boundary is somewhat lower than the typical values quoted in the literature primarily due to our definition: we consider the fully convective boundary to be the division between models that host a radiative core at some point in their lives, but that may be fully convective for the vast majority of their lifetimes, versus models that never have even a transient radiative zone.

\subsection{Checks and Physics}

To verify that this instability is visible in other stellar evolution codes apart from YREC, we made use of the publically available MESA \citep[][,http://mesa.sourceforge.net/]{paxton2011} stellar evolution code. We ran solar calibrated models with Z = 0.016498, Y = 0.26275, mixing length parameter $\alpha = 2.01$ and $Z/X = 0.02289$ \citep{grevesse1998} at 4.57 Gyr \citep{bahcall1995} and the default, out-of-the-box physics and numerical parameters released with MESA version 3372, with the exception of the model timesteps, which we forced to be of order $10^6$ years or smaller. We found the same qualitative behavior with \hethree{} at the convective boundary in the MESA models, with some quantitative differences. The fully convective boundary in these models occurs for masses $ M \le 0.281 \textrm{ M}_{\odot}$, and the mass range over which the convective episodes are present is $ 0.343 \textrm{ M}_{\odot} \le M \le 0.375 \textrm{ M}_{\odot}$ (versus $ 0.322 \textrm{ M}_{\odot} \le M \le 0.365 \textrm{ M}_{\odot}$ for the YREC models). We expect that this difference is due to the detailed differences between the input physics in the two different codes. Although the timescales differ somewhat, the behavior is qualitatively the same, which is evidence in favor of a physical phenomenon rather than numerical artifact. A MESA model for $M = 0.37 \textrm{ M}_{\odot}$ is shown in Figure \ref{fig1} among other YREC models for comparison. 

The exact masses at which the star is unstable to these convective episodes is also model dependent within YREC. For example, models run with \citet{kurucz1997} atmospheres rather than Allard atmospheres become unstable at $M = 0.332 \textrm{ M}_{\odot}$ rather than $0.322 \textrm{ M}_{\odot}$. However, in both cases the instability occurs over a mass range of $\Delta M = 0.043 \textrm{ M}_{\odot} $, consistent with a simple shift in the location of the fully convective boundary. Models run at non-solar metallicities also display unstable behavior over slightly different mass ranges. For example, a $0.34 \textrm{ M}_{\odot}$ model with $\textrm{[Z/X]} = -0.4$ (referenced to solar metallicity) displays instabilities with the longer timescales and larger amplitudes of a more massive $\sim 0.36$ solar metallicity object.

Physical processes such as convective overshoot have not been considered here, and will undoubtedly affect the exact location in mass of the instability boundaries. There are two distinct cases in which overshooting may affect the instability; the first is simply by shifting the effective location of the boundaries of the convection zones, resulting in a shift in the masses at which the instability occurs. The second effect is more subtle, and is related to the interplay and exchange of material between two closely separated convection zones due to overshooting.

The importance of overshooting in a system with closely separated convection zones has been considered in the case of A stars, which are thought to have two thin surface convection zones, one at the hydrogen ionization zone, and another at the deeper second ionization of He. These are formally separated by a radiative buffer, although it is unclear whether they are truly distinct from one another when overshooting is considered, as \citet{freytag1996, latour1981} suggest. In the particular case we consider here, overshooting would serve to connect the core and envelope convection zones either chemically or physically before they formally touch according to the Schwarzschild criterion. YREC treats overshooting as ``overmixing'' only, and does not adjust the temperature gradients in the overshoot regions, and so is incapable of capturing the full behavior of the system when the CZs are in this close configuration. One could imagine that the dilution of the core \hethree{} due to \hethree{} poor material mixed in through overshooting would lead to an overall drop in the \hethree{} luminosity in the core, a global decrease in the radiative temperature gradient, and therefore a shrinking of the convective core. Likewise, the mixing of \hethree{} rich material into the convective envelope might lead to increased burning, and cause the envelope convection zone to deepen slightly. Essentially, if the production timescale for \hethree{} in the core is much shorter than the timescale for the mixing of \hethree{} poor material into the central CZ, then the effects of overshooting should be minimal. If the mixing timescale is comparable or shorter than the production timescale for \hethree{}, then it is possible that the effects of overshooting could prevent these fully convective episodes from occurring at all. However, if the temperature gradients in the overshoot regions are very close to adiabatic, the boundaries of the the two CZs are actually closer in the case of overshooting than they would otherwise be. Further work on this interaction is therefore of interest. 

We offer, however, that the convective velocities in these stars should be low, and the stellar material strongly stratified. The extent of convective envelope overshooting in the Sun inferred from helioseismology is less than 0.05 pressure scale heights \citep{basu1997}, and both studies of cluster CMDs \citep{kozhurina1997,sarajedini1999,vandenberg2004} and asteroseismology \citep{briquet2011} suggest that the typical extent of core overshooting in intermediate mass stars is 0.1-0.2 pressure scale heights. There has been very little observational or theoretical investigation of the amount of overshooting in stars of a few tenths of a solar mass, but we might argue, given what information we have about other masses, that the degree of overshooting will be minimal. 

Likewise, we have neglected to include the effects of magnetic fields, which could be significant for very active M-dwarfs, and could potentially stabilize regions of the model against convection \citep[as in][]{mullan2001}.

\section{Observational implications} \label{sec:observations}

\subsection{Single Stars}
To investigate whether the luminosity and radius changes associated with the instability are observable, we focus on open clusters, in which we have a single, coeval, chemically homogeneous sample of stars. We chose a simple initial mass function, with $dn/dM = AM^{-\alpha}$, where $A$ is a normalization constant, and $\alpha = 1.3$ \citep{kroupa2001}. We sample the IMF in $0.001 \textrm{ M}_{\odot}$ bins, and determine our model luminosities and effective temperatures at fixed time for each mass. We add noise to the model HR diagram by assigning luminosities and effective temperatures randomly sampled from a normal distribution centered on the theoretical values for each stellar mass. We mimic the presence of binaries by randomly selecting half of the stars in the theoretical open cluster, and assigning a companion of mass $b M_{\textrm{primary}}$ where $b$ is a random number between 0 and 1 drawn from a uniform distribution, motivated by the distribution of binary mass ratios in \citet{raghavan2010}. We assign the binaries the luminosity weighted mean effective temperature of the two components, and add the two luminosities. We do not account for the effects of activity. The presence of the instability produces some weak structure along the single-star MS, which will be challenging to detect in any open cluster survey. However, a few stars lie below the MS, where, with the addition of proper motion or RV membership information, they might be recognized as 'odd' cluster members. Detection of such sources would rely on there being an open cluster with enough members, of an interesting age and metallicity such that this instability is visible. 

The presence of this instability could also manifest itself as a scatter in radius as a function of age in M dwarf wide binary systems in which the stellar parameters are well known. Given a sample of stars of the same mass and composition with a variety of ages, one should observe more scatter in the radii of the younger objects in comparison to the older objects. The effect will only be visible if one can disentangle the contributions from stellar activity and observational uncertainty, both of which are current challenges. It is important to note, however, that this is potentially an additional source of scatter in M-dwarf radii which is not related to activity or environment, but rather nuclear burning processes. 

\subsection{Cataclysmic variables}

Cataclysmic variables are short period binaries in which a white dwarf accretes mass through Roche lobe overflow from a relatively unevolved donor star. CVs are observed over a wide range of binary periods, ranging from $\sim 8$ hrs to just over 1 hour. The primary means of angular momentum loss are postulated to be magnetic braking associated with the stellar wind of the donor star, and losses through gravitational radiation, which operates most efficiently for the closest orbital separations. One of the most notable features in the CV period distribution is the so-called ``period gap'' \citep[see][]{rappaport1983, spruit1983} in which we observe a paucity of systems with periods of 2-3 hrs.

This instability is interesting in the context of CV systems. Given sufficient time and mass-loss rates, all CV donor stars will pass through the mass range over which these convective episodes occur in single-star models. The response of the donor star radius to mass-loss is a critical element of the CV picture; sudden changes could truncate the mass-loss by bringing the system out of Roche lobe contact, or conversely result in temporarily inflated accretion rates. Although a detailed discussion of the manner in which this instability interacts with the process of regulated mass-loss through Roche-lobe overflow is beyond the scope of this article, we do consider the response of single star models to constant mass-loss rates. Under the right conditions, this instability induces a significant, complicated response in the stellar radius, at an extremely interesting mass range in CV systems.

\begin{figure}
	\centerline{\includegraphics[scale = 1.0]{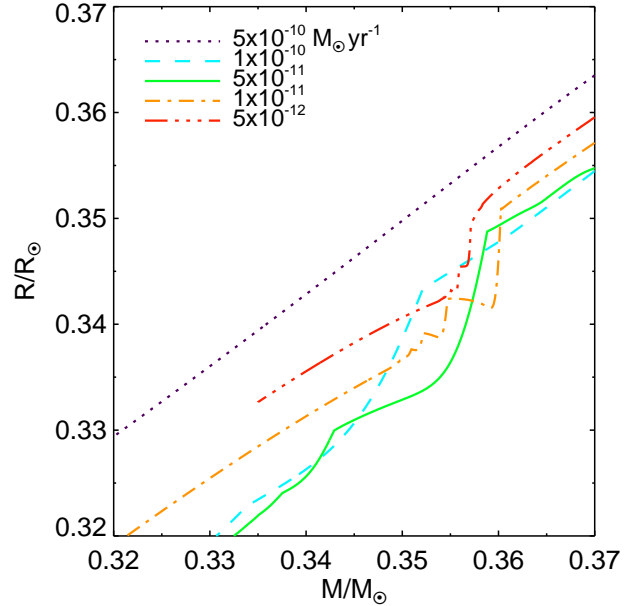}}
       \caption{Zoo of mass vs. radius curves for a sampling of mass-loss rates obtained using a starting model of $0.4 \textrm{ M}_{\odot}$ evolved to 1 Gyr before mass-loss is initiated. All mass-loss rates are in solar masses per year. The curves correspond to the following mass loss rates: dotted (purple): $5 \times 10^{-10} \textrm{ M}_{\odot} \textrm{ yr}^{-1}$, dashed (blue): $1 \times 10^{-10}$, solid (green): $5 \times 10^{-11}$, dot-dashed (orange): $1 \times 10^{-11}$, triple-dot-dashed (red): $5 \times 10^{-12}$. Note that the curve for  $5 \times 10^{-12}$ is truncated at 14 Gyr. }
	\label{fig3}
\end{figure}

 We apply constant mass-loss rates between $10^{-8}-10^{-12}  \textrm{ M}_{\odot} \textrm{yr}^{-1}$ to models of $0.4 \textrm{ M}_{\odot}$ and a single mass loss rate of $5 \times 10^{-11} \textrm{ M}_{\odot} \textrm{yr}^{-1}$ to models over a range of masses ($0.4-0.7 \textrm{ M}_{\odot}$), all of which have been evolved to 1 Gyr prior to beginning mass-loss. This exploration of of mass-loss rates is presented in Figure \ref{fig3}. In general, relatively unevolved models of the lowest masses ($\sim0.4 \textrm{ M}_{\odot}$) coupled with low mass-loss rates ($10^{-10} - 10^{-11} \textrm{ M}_{\odot} \textrm{yr}^{-1}$) show the largest radius ``glitches'' due to convective kissing.

Whether or not strong radius glitches appear in these models depends sensitively on a combination of the mass-loss rate, initial mass, and evolutionary state of the model. At fixed initial mass, the instability does not manifest itself in models until the mass-loss timescale is greater than the thermal timescale, $M/ \dot{M} \sim M^2/RL$. For high mass loss rates, the mass loss timescale is always smaller than the thermal timescale, the star is out of thermal balance, and no instability is present, even over the mass ranges where we expect it to occur. For lower mass-loss rates, the instability is present and the glitch in radius becomes visible (see Figure \ref{fig3}). 

The initial mass of the model sets the \hethree{} profile and abundance, which in turn determines if the \hethree{} abundance in the core is sufficient upon arrival at $\sim0.35 \textrm{ M}_{\odot}$ to become unstable. Low mass models have \hethree{} profiles peaked closer to the central region, and in general the absolute central \hethree{} abundance increases with decreasing mass for models of the same initial compositions. Higher mass models, which are more effective at burning \hethree{}, have \hethree{} profiles that peak at larger radii, and generally have more \hethree{} depleted cores. Finally, because models of different masses have different nuclear equilibrium timescales for \hethree{}, in general, the later the mass-loss begins, the less \hethree{} is present in the core of the model. These three effects conspire to make the radius glitches most visible in relatively unevolved, low mass models with low mass-loss rates. We note a small radius glitch occurs even in models with low mass-loss rates but insufficient \hethree{} to induce the instability, which occurs as \hethree{} is homogenized throughout the star as the model crosses the fully convective boundary; we believe this is the same feature noted in \citet{dantona1982} and \citet{andronov2004}.

Apart from the question of whether this instability affects the degree of mass loss through physically altering the radius of the donor star, one should also note that the composition of any donated material is influenced by the presence of the instability. Both \citet{shen2009} and \citet{shara1980} suggest that for \hethree{} mass fractions $\gtrsim 10^{-3}$  the ignition of novae may have some sensitivity to the \hethree{} abundance, and that the burning of accreted  \hethree{} on the primary can lead white dwarfs that are more luminous than their average accretion rates suggest. In the case of a 0.4 $\textrm{M}_{\odot}$ model with a constant mass loss rate of $5 \times 10^{-11} \textrm{M}_{\odot} \textrm{yr}^{-1}$, the surface \hethree{} abundance increases by a factor of 2 over the course of the first convective episode, from $\sim 5 \times 10^{-4}$ to $10^{-3}$. This increase in the \hethree{} abundance of accreted material would appear at masses above the fully convective boundary, where it would otherwise not exist. 

We have negelected to include rotational mixing in our models, which could be non-negligible in the case of such close binary systems in which the secondary may be relatively rapidly rotating due to tidal locking, which could help to drive meridional flows \citep{eddington1925, sweet1950, zahn1992} and result in mixing. Furthermore, mixing may also be driven by tidal distortions of the potential due to its close companion \citep{tassoul1982, andronov2004}. If the timescale for mixing is shorter than the timescale in between convective kissing episodes, the mixing of material could serve to eliminate episodic mixing events. With the inclusion of tidal deformation, the estimates of the minimum timescale for mixing are of order the timescale between convective episodes for systems at 3 hrs \citep[see][Fig. 9 for timescales]{andronov2004}. The initial feature in the curve in Figure \ref{fig3}, as the \hethree{} in the core is first mixed throughout the envelope would remain even in the case of strong mixing, but the subsequent variations in the radius would be damped. Therefore, the precise shape of these curves may be modified in the case in which rotational mixing is important, and our neglect of mixing is an important caveat to this discussion.

\section{Conclusion} \label{sec:conclusion}

We have discovered a novel instability in low mass, $M \sim 0.35 \textrm{ M}_{\odot}$, stellar models in which a buildup of \hethree{} in the core results in brief convective episodes within the first few Gyrs on the main sequence. These events are accompanied by changes in the radius and luminosity of the star. We have verified that the same instability appears to occur in the MESA code as well, which suggests that it is not a merely numerical artifact. Stars undergoing these convective episodes may be visible as both single stars or as members of binary systems. We have demonstrated that the radius of the donor star in a CV system could have a sizable response to the onset of the convective kissing instability, given donor stars in which the mass-loss rates, initial stellar mass, and evolutionary state ensures that there is sufficient \hethree{} available in the core by the time the stellar mass reaches and interesting range. The complicated structure of the glitches, which includes both increases and decreases in radius over the course of the instability, could interact in interesting ways with the process of CV mass-loss, either by severing contact, or by leading to temporarily inflated mass-loss rates, provided they are not suppressed by rotational mixing processes. We suggest that this instability be investigated in the context of full CV evolutionary models, where the interaction between the mass-loss rate and the convective kissing instability could be better quantified. 

\section*{Acknowledgments}
We would like to thank Lars Bildsten for useful discussion, and to acknowledge the KITP staff of UCSB for their warm hospitality during the research program ``Asteroseismology in the Space Age''. We also thank Ben Shappee and Andy Gould for useful discussions. This research was supported in part by the National Science Foundation of the United States under Grant No.\ NSF PHY05--51164, NSF Graduate Research Fellowship Grant RF\#743796 (JVS), and NASA grant NNX11AE04G (MHP). 

\bibliographystyle{apj}

\end{document}